Carbon-substitution dependent multiple superconducting gap of MgB$_2$: a "sub-meV" resolution photoemission study

S. Tsuda,[1] T. Yokoya,[1,*] T. Kiss,[2] T. Shimojima[1], S. Shin,[1,3] T. Togashi,[3] S. Watanabe,[1] C. Chen,[4] S. Lee,[5] H. Uchiyama,[5] S. Tajima,[5] N. Nakai,[6] and K. Machida[7]

[1] Institute for Solid State Physics, University of Tokyo, Kashiwa, Chiba 277-8581, Japan

[2] The Institute of Physical and Chemical Research (RIKEN), Wako, Saitama 351-0198, Japan

[3] The Institute of Physical and Chemical Research (RIKEN), Sayo-gun, Hyogo 679-5148, Japan

[4] Chinese Academy of Science, Beijing Center for Crystal R&D, Zhongguancuun, Beijing 100080, China

[5] Superconductivity Research Laboratory, ISTEC, Shinonome, Tokyo 135-0062, Japan

[6] Yukawa Institute, Kyoto University, Kyoto, Kyoto 606-8502, Japan

[7] Department of Physics, Okayama University, Okayama, Okayama 700-8530, Japan

*Present address: Japan Synchrotron Radiation Research Institute(JASRI)/SPring-8, Sayo-gun, Hyogo 679-5198, Japan

Abstract

"Sub-meV" resolution photoemission spectroscopy was used to study carbon-substitution dependence on the multiple superconducting gap of Mg(B$_{1-x}$C$_x$)$_2$. Two features corresponding to σ and π gaps are clearly observed in the raw spectra up to carbon concentration x = 7.5 %. The observed x dependence of the two gaps shows a qualitatively different behavior: a marked change of the σ gap proportional to the $T_c$ variation and a negligible one of the π gap. This as well as the temperature dependence can be explained with the two-band mean-field theory. Implications from the present study are discussed.

PACS: 74.70.Ad, 74.25.Jb, 79.60.-I

Superconductivity is a fundamental quantum phenomenon, which is now widely believed to be explained by the Bardeen-Cooper-Schreiffer (BCS) theory [1]. However, even after the appearance of the theory, continuous discoveries of new superconducting materials with higher transition temperatures ($T_c$) have made the field attractive owing to the expectations of a new theory as well as new materials having higher $T_c$. Therefore, the discovery [2] of the superconductivity on magnesium diboride ($MgB_2$) has also induced enormous studies to clarify the superconducting (SC) transition mechanism because it has the highest $T_c$ among inter metallic superconductors and has even higher $T_c$ than some of the cuprate superconductors with locating the $T_c$ close to the upper limit predicted by the BCS theory [1]. It is now generally accepted that $MgB_2$ has a multiple SC gap [3], originating in the two types of Fermi surface sheets with very different character, which simultaneously plays an important role for the high $T_c$. Theoretically, a possibility of the multiple SC gap was proposed more than 40 years ago in terms of "two-band superconductivity". However, further experimental verification is essential for complete understanding of this unique superconductor. Here, we show carbon-substitution dependence of the two gap of $MgB_2$ with the "sub-meV" resolution photoemission spectroscopy (PES). The result clearly shows two gaps and different substitution dependence for the two gaps. These results shed another light on to the two-band superconductivity in $MgB_2$, and also suggest a new road to explore the higher $T_c$ material.

MgB$_2$ has two layers consisted of boron and magnesium respectively [2]. The first-principles calculations [4,5] have indicated that MgB$_2$ is a band metal, where four bands are crossing Fermi level ($E_F$). Two of them originate in the boron 2p $\sigma$ orbital with a two-dimensional character, named $\sigma$ band, and the other two do in the boron 2p $\pi$ one with a three-dimensional character, named $\pi$ band. The calculated band structure is consistent with experimental one, *e.g.* obtained from the angle-resolved photoemission spectroscopy [6,7,8]. At this stage, the transition mechanism is explained by the electron-phonon coupling [9,10], where the coupling strength depends on the symmetry of the Fermi surface. This scenario, based on the two-band model [9,11] or, in more detail, *k*-dependent BCS theory or known as Eliashberg theory [10,12], has succeeded in explaining the multiplicity of the SC gap qualitatively and the $T_c$ quantitatively. Conventional BCS theory [1] predicts that a reduction of the density of states (DOS) at $E_F$ reduces $T_c$ and a SC gap value. For the MgB$_2$ case, the electron doping is supposed to decrease $T_c$ because of a decrease of the total DOS at $E_F$: the $\sigma$ band at $E_F$ will be occupied and the $\sigma$-band DOS are reduced by the doped electrons while the DOS at $E_F$ of the $\pi$ band will be almost unchanged [5]. Actually the electron-doped MgB$_2$ by chemical substitution of boron with carbon shows lower $T_c$ than pure MgB$_2$ [13]. Although a SC gap size has important information to estimate the stability of the SC state, systematic direct observation on the SC gap of chemical substituted MgB$_2$ is not sufficient. For this purpose, ultrahigh-resolution photoemission spectroscopy is the best method because it can reveal the multiple SC gap directly, as was reported with PES using a He discharging lamp [14]. Moreover, recent advance in energy resolution of a photoemission spectrometer by using a laser as a photon source has achieved a "sub-meV" resolution [15] and provided us new opportunity to do more quantitative and reliable study for the multiple gap and for the role of the interband coupling.

Photoemission measurements were performed on a spectrometer built using a Scienta R4000 electron analyzer and a ultra-violet laser (6.994 eV). The energy resolution was set to ~ 0.4 meV. Samples are cooled using a flowing liquid He refrigerator with improved thermal shielding. The pressure of the spectrometer was better than 1 x $10^{-10}$ Torr. during all the measurements. All the photoemission measurements have been done for *in-situ* fractured surfaces. Temperature-dependent spectral changes were confirmed by cycling temperature across $T_c$. $E_F$ of samples for high-resolution measurements was referenced to that of a gold film evaporated onto the sample substrate and its accuracy is estimated to be better than ± 0.2 meV.

Here we selected carbon as a substituting material to control $T_c$ or, in other words, to modulate the superconductivity. Carbon is the only atom which can be replaced by boron, giving us a unique opportunity to study changes in boron electronic states which plays a crucial role for the superconductivity. Moreover, the systematic carbon concentration (x) dependent studies on $T_c$, the axis length *et al*. using single crystals are already available [16]. The carbon substituted $Mg(B_{1-x}C_x)_2$ polycrystalline samples were synthesized under the high-temperature and the high-pressure conditions [16]. The carbon concentrations (x) and $T_c$s used in this study are listed in table 1.

Figure 1 shows the "sub-meV" resolution photoemission spectra as a function of x obtained at 3.5 K (superconducting state). In the spectrum of x = 0 %, we clearly observe two well-separated peaks at 2.6 and 7.1 meV binding energy, which are attributed to SC gaps on π- and σ- band. Compared with the spectrum obtained using a conventional He discharge lump (indicated by a black thin line), the quality of the spectrum was drastically improved and the multiplicity of the SC gap is certain because of the marked increase of the energy resolution. Moreover, no photoemission intensity was observed around $E_F$, directly indicating that $MgB_2$ is a full gap superconductor. As the x is increased, while the higher binding energy peak shifts clearly, the lower binding energy structure stays nearly the same position. Within the measured x concentrations, the two structures do not merge, indicating that $Mg(B_{1-x}C_x)_2$ has two gaps up to x = 7.5 %, consistently with the magnetic torque experiment [17]. These observations, which clearly identify the carbon-substituted change in the SC gap from the raw data alone, are very important, demonstrating the advantage of "sub-meV" resolution PES.

To quantitatively estimate the SC gap size, we tried to fit the spectra using Dynes function [18,19]. Figure 2 shows the x dependence of the gap size at T = 0 K. The gap sizes at T = 0 K were determined by an extrapolation from the fitting result using a known BCS temperature dependence of a SC gap [1]. The larger gap, corresponding to the σ band, decreases with increasing x, while the smaller gap, corresponding to the π band, looks independent to x within the fitting error up to x = 7.5 %. Bussmann-Holder and Bianconni calculated the x dependence of the two gaps for Al substituted case [20], which predicted that the smaller gap increases up to the 20 % substitution. The result is different from the present one. The x dependence of $T_c$ shown on fig. 2 by a solid line seems very similar to that of the larger gap in contrast to that of the smaller gap. The similar x dependence of the larger gap and $T_c$ means the reduced gap $2\Delta/k_BT_c$, which is a measure of coupling strength between the SC electron, is independent to x. In the weak coupling limit, this value is 3.52, which is known as the mean field BCS value [1]. In the present case for the larger gap, this value is 4.1, which is surely larger than the BCS value, and it is almost independent to x. This result suggests that electron-phonon coupling of the σ band is strong and also $T_c$, or, in other words, superconductivity is governed by the σ band.

To get further insight into the nature of inter- and intraband couplings, it is useful to compare the experimental result with a model, which is as simple as possible. Here we use the modified BCS theory extended to multi band discussed by H. Suhl *et al*. [11], which can deal with the multi-gap superconductivity. This model contains four parameters (the electron-phonon coupling matrix element) producing the two intraband couplings for the larger and smaller gaps, ($\lambda_{\sigma\sigma}$, $\lambda_{\pi\pi}$), and the two interband couplings between the larger gap and the smaller gap ($\lambda_{\pi\sigma}$, $\lambda_{\sigma\pi}$) after being multiplied by corresponding calculated partial DOS at $E_F$ [4,5]. In Fig. 2, we plotted the x-dependent and *T*-dependent two gap values (marks), respectively, compared with calculated curves using the only four coupling constants plotted in Fig. 3. We found that this simple model can reproduce the gap sizes for both the gaps at $T = 0$ and their temperature dependence very well [21]. From Fig. 3, it is clear that the intraband coupling of the σ band decreases while the interband coupling increases with increasing x. The suppression of the $\lambda_{\sigma\sigma}$ cannot be explained by the DOS at $E_F$ of the σ band alone, since the decrease of the calculated DOS at $E_F$ of the σ band due to electron doping [4,5] is estimated to be, at most, 10 % for x = 7.5% within the rigid band model, much smaller than the 20 % reduction in $\lambda_{\sigma\sigma}$. This suggests that a suppression of the electron-phonon coupling, as discussed from Raman studies [22], and/or an increase of interband impurity scattering are also responsible for the reduction of $T_c$, though theoretical studies have reported negligible interband impurity scattering effects for the carbon substitution [23].

Thus we made firm experimental confirmation on the two-band superconductivity of $MgB_2$. Lastly, we would like to comment on an implication for $T_c$ from the two-band model. The multi band model gives $T_c$ by

$$T_c = \Theta_D \exp(-1/\lambda),$$

where $\Theta_D$ is the Debye temperature and $\lambda$ is the total coupling constant given by

$$\lambda = \frac{\lambda_{\sigma\sigma} + \lambda_{\pi\pi}}{2} + \sqrt{\left(\frac{\lambda_{\sigma\sigma} - \lambda_{\pi\pi}}{2}\right)^2 + \lambda_{\sigma\pi}\lambda_{\pi\sigma}} \ .$$

If the interband coupling constants are zero, $T_c$ is completely governed by $\lambda_{\sigma\sigma}$. However,

if the interband coupling is not zero, $T_c$ will always become larger than the case of $\lambda_{\sigma\pi}\lambda_{\pi\sigma} = 0$. Thus the role of the interband coupling is to make $T_c$ higher than the case without interband coupling. In the present case, using obtained parameters, $T_c$ is estimated to be pushed up about 5 K by the interband coupling. Normally, it is believed that a larger electron-phonon coupling necessary for an increase of $T_c$ gives rise to another transition making the system insulating. In $MgB_2$, the existence of the $\pi$ band may play some role for preventing such an instability occurring at the $\sigma$ band with the strong electron-phonon coupling and allow $\sigma$ band to become a strong coupling superconductivity at higher $T_c$. Moreover, the appropriate interband coupling also helps the high $T_c$. Thus searching a new two-band superconductor is meaningful and hopeful to obtain a much higher $T_c$.

In conclusion, the present study provides a direct observation of the multiple gap and its carbon substitution dependence using the "sub-meV" PES. In addition, it provides a qualitative estimate of the substituted dependence of inter- and intraband coupling between the $\sigma$ and the $\pi$ band using the simple two-band model supposing the multi-band mean field approximation, giving strong verification for the two band superconductivity in $MgB_2$.

This work was supported by Grant-in-aid from the Ministry of Education, Science, and Culture of Japan. This work was supported by the New Energy and Industrial Technology Development Organization (NEDO) as Collaborative Research and Development of Fundamental Technologies for Superconductivity Applications. S. T. and H. U. thank to JSPS fellow ship for financial support.

Captions

FIG. 1. (color) "Sub-meV" resolution photoemission spectra of carbon substituted $MgB_2$ polycrystalline samples for the carbon concentrations x = 0%, 2%, 5%, and 7.5%. The black thin line indicates the photoemission spectrum of a pure $MgB_2$ polycrystalline

sample obtained using a conventional He discharge lump with the energy resolution of ~ 3.0 meV. Each vertical short bar is a guide to the eyes, corresponding to the position of the superconducting gap, and hatched areas emphasize two gaps. The larger gap corresponds to the gap of the σ band and the smaller gap corresponds to the gap of the π band.

FIG. 2. (color) (a) Carbon concentration dependence of the superconducting gap for the larger and smaller gaps. The carbon concentration dependence of the transition temperature is indicated by a solid line (right axis). The experimental results are compared with model calculation results indicated by broken lines [11]. The experimental results are well reproduced by the calculations. The larger gap and the $T_c$ show very similar carbon concentration dependence up to 7.5 %. (b) Temperature dependence of the superconducting gaps for each carbon concentration (marks) determined using the Dynes function analysis. The vertical axis is normalized by the larger gap size of x = 0% and the horizontal axis is normalized by the $T_c$ of x = 0 %. The calculated result indicated by solid curves reproduces the experimental result well.

FIG. 3. (color) Carbon concentration dependence of the coupling parameters. It is clear that the intraband coupling of the σ band decreases while the interband coupling increases with increasing carbon concentration.

TABLE I. Relation between the carbon concentration and the $T_c$.

| Carbon concentration (%) | $T_c$ (K) |
| --- | --- |
| 0 | 38 |
| 2 | 35.5 |
| 5 | 30 |
| 7.5 | 24 |

Fig. 1

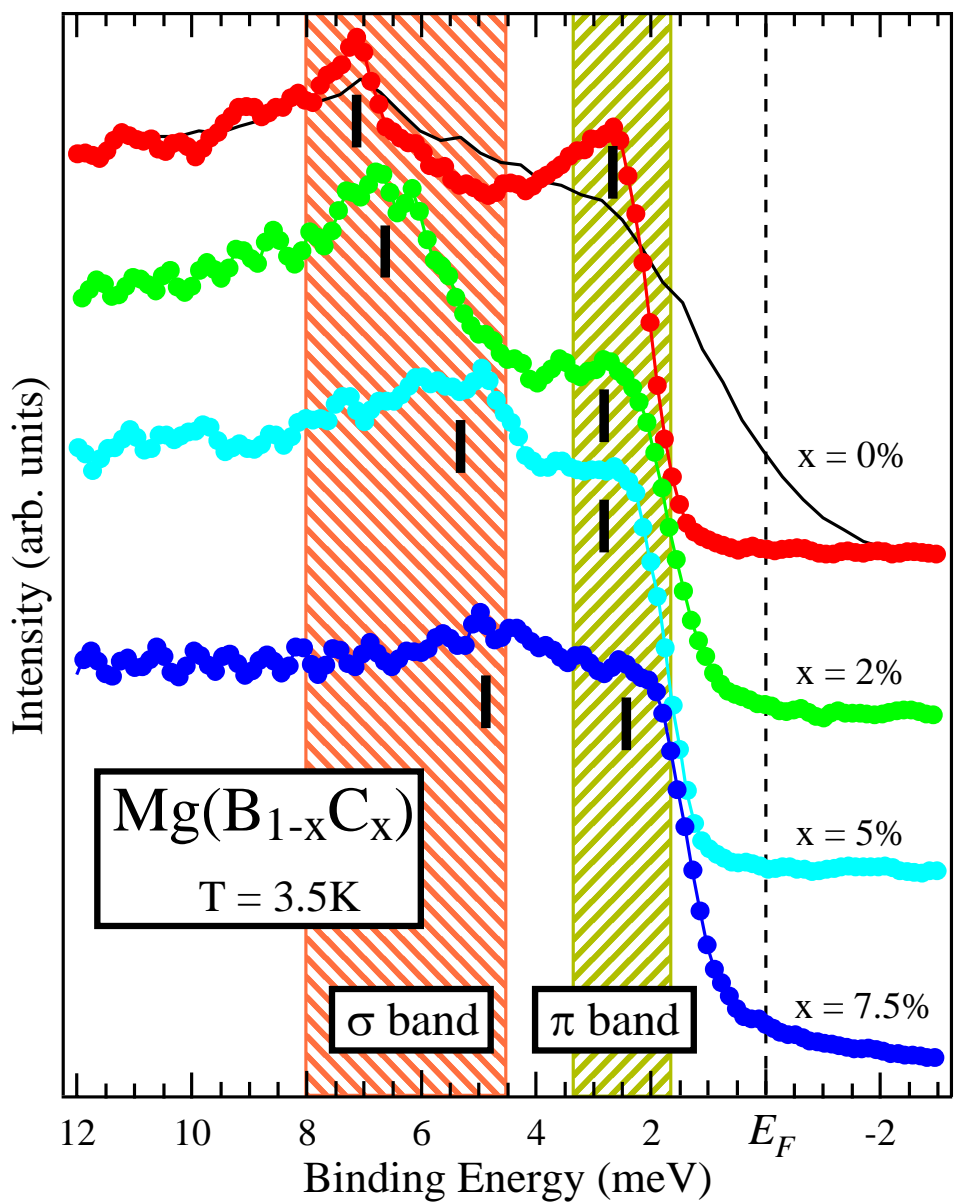

Fig. 2

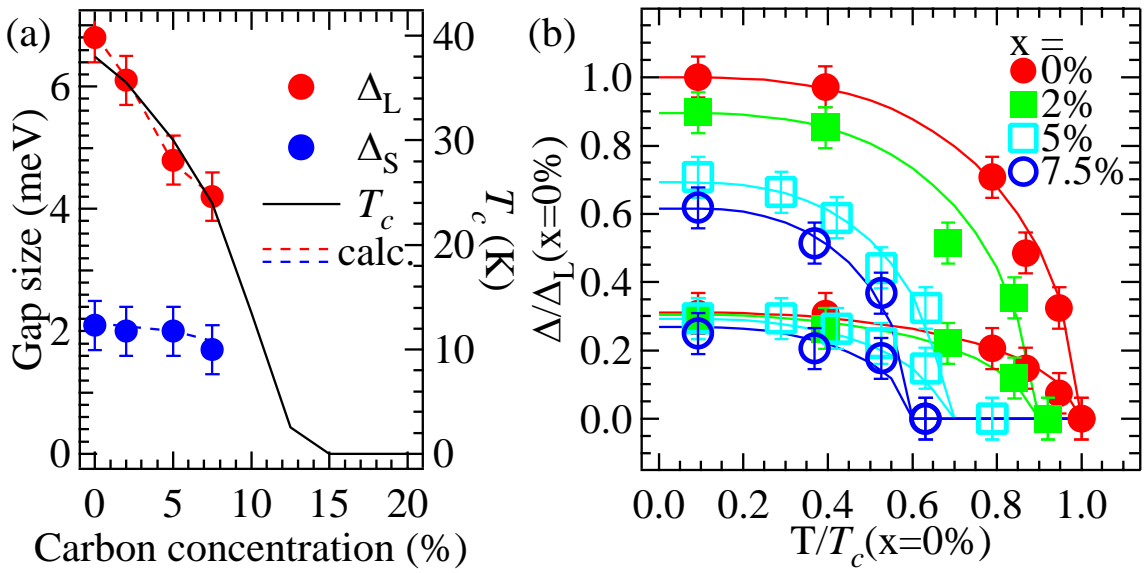

Fig. 3

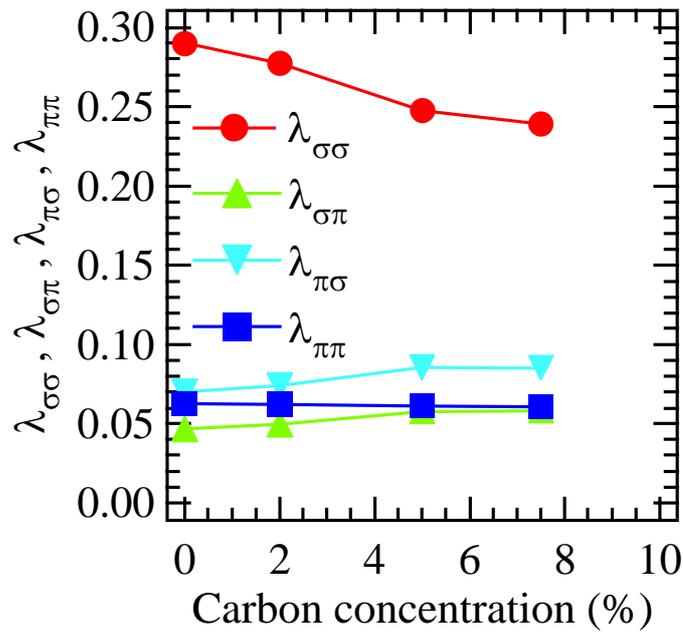